\documentclass[12pt]{iopart}

\usepackage{graphicx}
\usepackage{subfigure}

\usepackage{iopams}  
\begin{document}

\title{Relativistic Quantum Dynamics on a Double Cone}

\author{F A Gomes$^1$, E O Silva$^2$, Jonas R F Lima$^3$,C Filgueiras$^4$, F Moraes$^1$,}

\address{$^1$Departamento de F\'{\i}sica, CCEN,  Universidade Federal da Para\'{\i}ba, Caixa Postal 5008,  58051--900, Jo\~ao Pessoa, PB, Brazil}
\address{$^2$Departamento de F\'{i}sica, Universidade Federal do Maranh\~{a}o,
  Campus Universit\'{a}rio do Bacanga, 65080--805, S\~{a}o Lu\'{i}s, Maranh\~{a}o, Brazil}
\address{$^3$Departamento de F\'{\i}sica, Universidade Federal Rural de Pernambuco, 52171--900, Recife, PE, Brazil}
\address{$^4$Departamento de Física (DFI),
Universidade Federal de Lavras (UFLA), Caixa Postal 3037,
37200--000, Lavras, Minas Gerais, Brazil}

\ead{fagfisica@gmail.com (Felipe A Gomes), edilbertoo@gmail.com (Edilberto O Silva), jonas.iasd@gmail.com (Jonas R F Lima),
cleversonfilgueiras@yahoo.com.br (Cleverson Filgueiras), moraes@fisica.ufpb.br (Fernando Moraes)}

\date{\today}

\begin{abstract}
In this paper, we study the relativistic quantum problem of a particle constrained to a double cone surface. For this purpose, we build the Dirac equation in a curved space using the tetrads formalism. Two cases are analysed. First, we consider a free particle on a double cone surface, and then we add an uniform magnetic field. The energy spectrum is obtained and the instability of the motion is discussed. We show that the magnetic field breaks the nappe degeneracy, inducing different energy spectra for each nappe. The results obtained here can be applied, for instance, in the investigation of the electronic and transport properties of condensed matter systems that can be described by an effective Dirac equation, such as graphene and topological insulators.
\end{abstract}

\vspace{2pc}
\submitto{\JPA}
%\maketitle

\section{Introduction}

The dynamics of quantum particles on  surfaces is an exciting research area due to its experimental realization with interfaces, like in the Quantum Hall Effect \cite{klitzing}, or real two--dimensional materials like graphene \cite{novoselov2004electric} and its curved versions (nanocones, nanotubes, etc.) and topological insulators \cite{qi2010quantum}.  Conical surfaces, in particular, are of special interest in the study of quantum phenomena due to the singular curvature they carry \cite{filgueiras2008quantum}. Effectively, they appear in investigations on topological defects in continuum media \cite{furtado1999landau}, cosmic strings \cite{bakke2010bound} and even black holes \cite{furtado1997electrostatic}.  In recent years, the quantum dynamics of charge carriers in the presence of different topological defects have been investigated in the aforementioned materials, such as dislocations \cite{schmeltzer2012geometrical, ran2009one} and disclinations \cite{lima2014effects, bueno2012landau}.

Double conical surfaces have been much less studied. The classical and quantum  dynamics of a particle moving on a double cone was presented in Ref. \cite{kowalski2013dynamics}.  Classically, the apex of the cone works like a filter, allowing only particles with zero angular momentum to pass from one cone to the other, following a straight line. Consequently, any small perturbation in the angular momentum causes a drastic change in the trajectory of the particle, making the rectilinear motion unstable.  Traces of instability in the movement are also observed in the quantum regime \cite{kowalski2013dynamics}.

In Ref. \cite{lopes2015theoretical}, \textit{ab initio} calculations were made with carbon double cones. The results show that they have low formation energy  and therefore it is expected that they may be obtained experimentally. This possibility motivated us to extend the work of Ref. \cite{kowalski2013dynamics} to the relativistic domain, looking specifically at the Dirac equation and its solutions for a particle on the double cone both with and without a magnetic field. Since a class of low--dimensional materials \cite{qi2010quantum, bueno2012landau, jonasjap, vozmediano2010gauge} have been theoretically described by an effective Dirac equation, the results obtained here may be applied to condensed matter systems as well.

The organization of the paper is as follows. In Sec. \ref{2} we describe the geometric model for a double cone surface. We explain in this section why the usual coordinate system used for a conical surface with a single nappe is inconvenient for a double cone surface, and build a new coordinate system extending the radial coordinate.  Using this reference frame we obtain the solution of the Dirac equation for the free particle on the conical surface in Sec. \ref{3}. We analyse the behavior of the solution for angular momentum number near $0$ and find "scars" of instability, like those in Ref. \cite{kowalski2013dynamics}. In Sec. \ref{4}, the problem is extended  to the case which includes an external uniform magnetic field parallel to the cone axis. There, we obtain the energy spectrum and analyse how the geometry of the surface changes the spectrum. The paper is summarized and concluded in Sec. \ref{5}.

\section{Geometric approach}
\label{2}

\begin{figure}[hpt]%[hbtp]
\centering
\includegraphics[width=9cm,height=8cm]{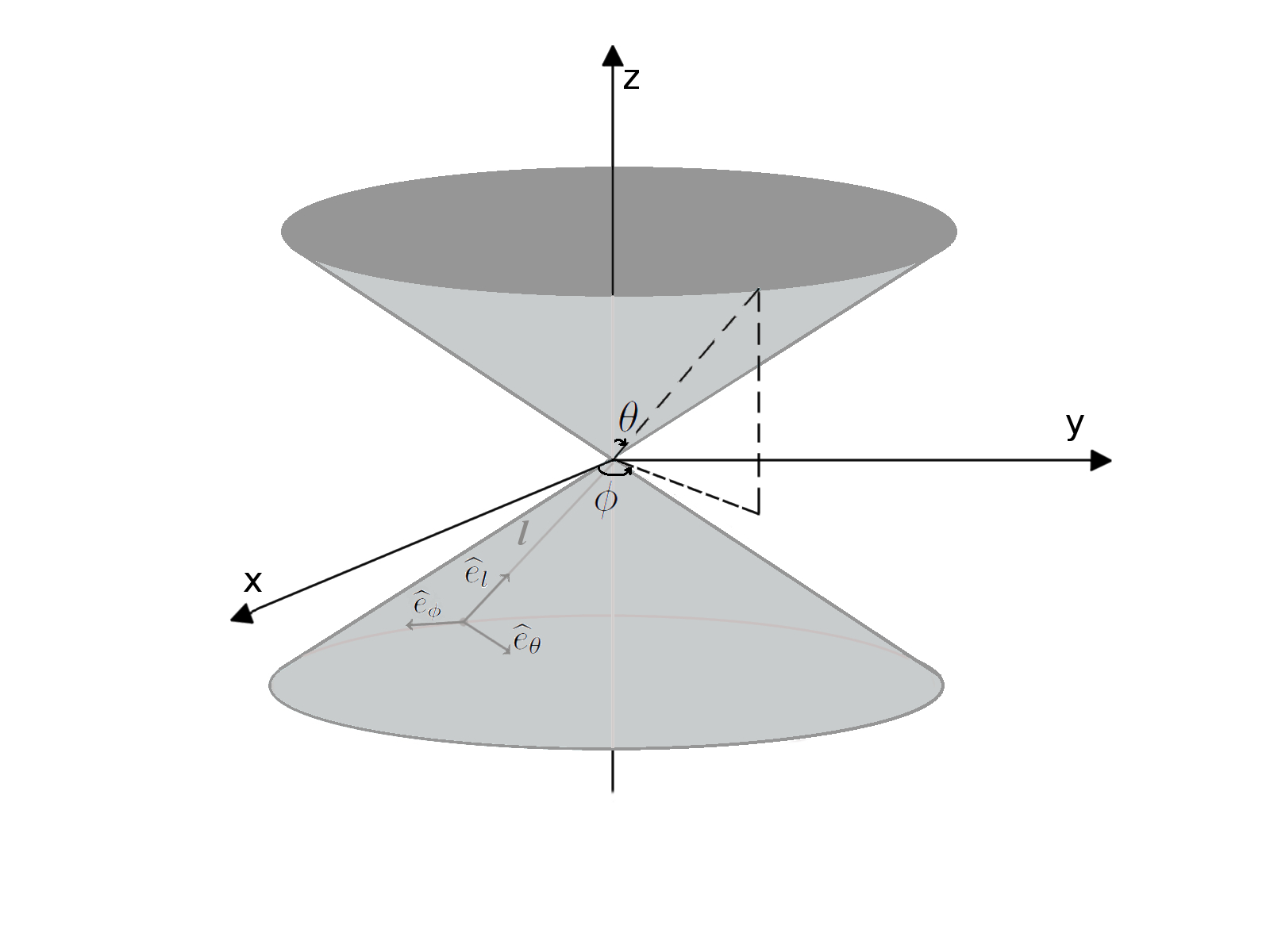}	
\caption{Coordinate system for a double cone surface. With the extension of the radial coordinate domain, a point in the bottom nappe has the same $\theta$ coordinate as a point in the upper nappe.}
\label{figure1}
\end{figure}

Considering the spherical coordinate system, we can construct the surface of a cone simply by fixing the angular coordinate $\theta$. In the case of a double cone surface, a drawback comes up when we use this approach. Since a double cone has two nappes, we can not use the spherical coordinate system with just a fixed value for the angular coordinate $\theta$, since it is necessary two values, one for each nappe. In order to avoid the complications of working with a discrete coordinate (for instance: to establish a metric in this space would be a problem),  we extend the radial coordinate domain for the entire set of real numbers, as done in \cite{kowalski2013dynamics}. Accordingly, the bottom nappe corresponds to the negative values of the radial coordinate, but with the same value of $\theta$ as the points on the upper nappe (see Fig. \ref{figure1}).

The coordinate system used in this article can be described by the following relations \cite{kowalski2013dynamics}:
\begin{eqnarray}
	x  &=& l \sin\theta \cos\phi ,\nonumber \\
	y  &=& l \sin\theta \sin\phi ,\\
	z  &=& l \cos\theta ,\nonumber
\label{coord1}
\end{eqnarray}
with the respective domains given by
\begin{eqnarray}
l &\in & (-\infty,+\infty) ,\nonumber \\
\theta &\in & \left( 0,\frac{\pi}{2} \right) ,\\
\phi &\in & [0,2 \pi) . \nonumber
\label{domain}
\end{eqnarray}
The equation of the double cone surface in terms of this coordinate system is therefore $\theta= const$,  whereas in the three--dimensional Cartesian coordinates it is $x^2 + y^2 = z^2\tan^2 \theta$. The induced metric of the cone is thus $ds^2=dl^2 + l^2\sin^2\theta  d\phi^2$.  In the $2+1$ dimensional spacetime, following the signature convention $(-++)$, the line element is then given by
\begin{equation}
ds^2=-c^2dt^2+dl^2+l^2\alpha^2d\phi^2 ,
\label{metric}
\end{equation}
where $\alpha= \sin \theta$ and therefore can assume values only in the $0<\alpha<1$ range.

We are interested in investigating the relativistic problem of a particle constrained to a double cone surface and searching for instabilities in its motion, in the spirit of the nonrelativistic problem studied in Ref. \cite{kowalski2013dynamics}. In the following section, we  build the Dirac equation for a free particle   within the geometric scenario discussed above.

\section{Dirac equation of a free particle on a double cone surface} 
\label{3}

In order to build the Dirac equation on a curved space,  one can use the tetrad formalism \cite{birrell1984quantum, misner1973gravitation}. This formalism is required in a curved space because in this background the spinors must be defined locally. A local reference frame can be built by a noncoordinate basis $\widehat{\theta}^a = e^a_\mu(x) dx^{\mu}$, whose components $e^a_\mu(x)$ are called tetrads. As notation, we are using Greek indices for coordinates of the curved spacetime, while Latin indices denote the local reference frame of the observers.

We can use the tetrads formalism to relate the metric for the curved space to the  flat space metric
\begin{equation}
g_{\mu\nu}(x)=e^a_\mu(x) e^b_\nu(x) \eta_{ab} ,
\label{metricflattocurved}
\end{equation}
where $\eta_{ab}=diag\left(-c^2, +1, +1\right)$ is the Minkowski metric tensor. In order to write the Dirac equation for a curved background, we use the tetrads $e^a_\mu(x)$ to intermediate some changes in the usual equation. The gamma matrix, $\gamma ^\mu$, will be now defined in terms of the field $e^a_\mu(x)$ and of the standard flat space Dirac matrices $\gamma ^a$,
\begin{equation}
\gamma ^\mu (x)=e^\mu _a (x) \gamma^a .
\label{gammacurved}
\end{equation}
The covariant derivative of spinor field is given by $\nabla_\mu \Psi=\partial _\mu \Psi + \Omega _\mu \Psi$, where the spinorial connection is
\begin{equation}
\Omega _\mu=\frac{i}{4}\omega _{\mu ab}\Sigma^{ab} ,
\label{spinorialconnection}
\end{equation}
 $\Sigma^{ab}=\frac{i}{2}[\gamma^a,\gamma^b]$ and $\omega _{\mu ab}$ is a 3--form, known as spin connection. With these elements, we can write the Dirac equation for a curved space as
\begin{equation}
[i \hbar c \gamma^\mu (\partial _\mu + \Omega _\mu ) - mc^2]\Psi=0 .
\label{diraccurved}
\end{equation}
For a double cone surface, which has the metric (\ref{metric}), the condition (\ref{metricflattocurved}) allows us to choose the triad
\begin{equation}
e^\mu _a = \left(\begin{array}{ccc}
1 & 0 & 0 \\
0 & 1 & 0 \\
0 & 0 & \frac{1}{\alpha l} \end{array} \right) .
\label{gammacurved2}
\end{equation}
The spin connection is obtained from the first Cartan structure equation, given by $d\hat{\theta}^a+\omega^a_b \wedge \hat{\theta}^b=0$. Since $\omega^a_b=\omega^a _{\mu b} dx^\mu$, we get the following nonnull spin connection components
\begin{equation}
\omega_{\phi21}=-\omega_{\phi12}=\alpha .
\label{spinconnection}
\end{equation}
The relation between the spin connection and the spinorial connection is expressed in (\ref{spinorialconnection}). Using this relation and the result (\ref{spinconnection}), we obtain only one nonnull component for the spinorial connection
\begin{equation}
\Omega _\phi = -\frac{i}{2}\alpha\Sigma^3 .
\label{spinorialconn}
\end{equation}

Now, we need to determinate the gamma matrices for a double cone surface. For this purpose, we use the relations (\ref{gammacurved}) and (\ref{gammacurved2}) and obtain that
\begin{eqnarray}
&\gamma^t & = \gamma^0 ,\nonumber \\
&\gamma^l & = \gamma^1 ,\nonumber \\
&\gamma^\phi & = \frac{\gamma^2}{\alpha l} .
\label{gammas}
\end{eqnarray}
With all these quantities determined for the double cone surface, we can write the resulting Dirac equation as
\begin{eqnarray}
 i \hbar c  \left[ \frac{\gamma^0}{c}\partial _t  +  \gamma^1  \partial_l + \frac{\gamma^2}{\alpha l} \left( \partial _\phi -\frac{i}{2}\alpha\Sigma^3 \right) \right]\Psi - mc^2  \Psi=0 .
\label{free1}
\end{eqnarray}
It is natural to choose as \textit{ansatz}, solutions like
\begin{equation}
\Psi(t,l,\phi)=e^{-\frac{iEt}{\hbar}}e^{ij \phi} \psi_l ,
\label{ansatz}
\end{equation}
where $E$ and $j$ are constants of integration and can be interpreted as energy and angular momentum quantum number, respectively. Periodicity in the azimulthal angle $\phi$ implies that $j = 0, \pm1, \pm2, \pm3 ..., j \in Z$. With the \textit{ansatz} (\ref{ansatz}), we write the equation (\ref{free1}) in terms of two coupled differential equations
\begin{eqnarray}
\left(E- mc^2  \right)\psi_A  = -i\hbar c \sigma^1 \left(\partial _l + \frac{1}{2l}\right)\psi _B + \frac{\hbar c}{\alpha l} \sigma^2 j \psi _B,
\label{coupled1}
\end{eqnarray}
and
\begin{eqnarray}
\left(E+ mc^2 \right)\psi_B=-i\hbar c \sigma^1 \left(\partial _l + \frac{1}{2l}\right)\psi _A + \frac{\hbar c}{\alpha l} \sigma^2 j \psi _A ,
\label{coupled2}
\end{eqnarray}
where  $\psi _l = \left(\begin{array}{c}
\psi_A  \\
\psi_B  \\
 \end{array} \right)$. Uncoupling these equations,  we get
\begin{eqnarray}
l^2\partial ^2_l \psi_A + l \partial _l \psi_A - \left[ \frac{j^2}{\alpha^2} - \frac{j \sigma^3}{\alpha} + \frac{1}{4} \right] \psi_A + \frac{l^2}{\hbar^2c^2} \left( E^2- m^2c^4 \right) \psi_A = 0 ,
\label{differential11}
\end{eqnarray}
\begin{eqnarray}
l^2\partial ^2_l \psi_B + l \partial _l \psi_B - \left[ \frac{j^2}{\alpha^2} + \frac{j \sigma^3}{\alpha} + \frac{1}{4} \right] \psi_B + \frac{l^2}{\hbar^2c^2} \left( E^2- m^2c^4 \right) \psi_B = 0 ,
\label{differential12}
\end{eqnarray}
As the only matrices in the equations (\ref{differential11}) and (\ref{differential12}) are $\sigma^3$ and the identity, and both obviously commute, we can diagonalize them simultaneously, and therefore, with $\sigma^3 \psi_A= s \psi_A$ and $\sigma^3 \psi_B = s \psi_B$, where $s = \pm1$ are the eigenvalues of $\sigma^3$ corresponding to spin up and spin down, we write the above equations as: 
\begin{eqnarray}
l^2\partial ^2_l \psi_A + l \partial _l \psi_A + \left[ K^2 l^2 - \nu_A^2 \right] \psi_A = 0 ,
\label{differential1}
\end{eqnarray}
\begin{eqnarray}
l^2\partial ^2_l \psi_B + l \partial _l \psi_B + \left[ K^2 l^2 - \nu_B^2 \right] \psi_B = 0 ,
\label{differential2}
\end{eqnarray}
where we defined
\begin{eqnarray}
 &K^2&=\frac{1}{\hbar^2 c^2} \left(E^2 - m^2c^4 \right), \qquad \nu_A^2 = \left( \frac{j^2}{\alpha^2}- \frac{sj}{\alpha}  + \frac{1}{4} \right), \qquad \\\nonumber
  &\nu_B^2 & = \left( \frac{j^2}{\alpha^2}+ \frac{sj}{\alpha}  + \frac{1}{4} \right).
\label{nuek}
\end{eqnarray}
The expressions (\ref{differential1}) and (\ref{differential2}) are Bessel equations. We can write the solution of these equations as a combination of Bessel functions of the first kind. We avoid the solution of second kind because it diverges at  the origin of the coordinate system.

In order to compare our results with the solution of the Schr\"odinger equation obtained in Ref. \cite{kowalski2013dynamics}, let us consider the nonrelativist limit of the Dirac equation for the double cone surface. This limit is obtained by assuming $E = mc^2$ and defining the nonrelativistic energy as the energy measured from $mc^2$, that is, $\epsilon = E - mc^2$. In that limit the component $\psi_A$ is much bigger that $\psi_B$, so we neglect $\psi_B$. For $\psi_A$ we get
\begin{equation}
l^2\partial ^2_l \psi_A + l \partial _l \psi_A + \left[ \frac{2m\epsilon}{\hbar^2} l^2 - \nu_A^2 \right] \psi_A = 0 ,
\label{limit}
\end{equation}
which is basically the same equation obtained in Ref. \cite{kowalski2013dynamics}, except by the spin-orbit term $\left(\frac{sj}{\alpha} \right)$ presents in $\nu_A$. Although this correction term, the conclusions about instability in Ref. \cite{kowalski2013dynamics} remains the same. The Sch\"odinger equation was obtained in the aforementioned article by using a linear momentum operator modified to become self--adjoint. We do not use this approach in the relativistic case because all the information on the geometry is already incorpored into the Dirac equation through the metric and the spinorial connection. This takes care of the self--adjointness of the operators involved.

As discussed in Ref. \cite{kowalski2013dynamics}, the dynamics of a particle in a double cone is intimately related to its angular momentum. In the classical approach, using the Hamiltonian formalism, it is shown that the cone apex, that is, the origin of the coordinate system, works like a filter and only particles with zero angular momentum can travel between the two cones. When the quantum dynamics is analyzed, it is shown that traces of classical instability remain in the quantum regime. The instability in the quantum regime is due to the closure relation for the Bessel functions. This relation says that the solutions $J_{\nu}(x)$, with $\nu <-\frac{1}{2}$, can not be normalized and hence do not represent physically acceptable solutions. This condition leads to an abrupt change in the behavior of the wave function near $ \nu=-\frac{1}{2}$. Inspired by this behavior in the nonrelativistic case, we check what happens with the relativistic quantum dynamics of a particle in the double cone surface.

Let us analyse the solution of (\ref{differential1}) and (\ref{differential2}) with $j=0$ and $j \rightarrow 0$. For $j=0$ we have 
\begin{equation}
\psi_{A}= \frac{1}{\sqrt{l}}\left[A_1 \sin(Kl) + A_2 cos(Kl)\right],
\label{psiA.2}
\end{equation}
\begin{equation}
\psi_{B}= \frac{1}{\sqrt{l}}\left[B_1 \sin(Kl) + B_2 cos(Kl)\right] .\label{psiB.2}
\end{equation}
For $j \neq 0$ we can express the solution as a linear combination of Bessel functions with positive and negative indices
\begin{equation}
\psi_{A}= A_1 J_{\sqrt{\left( \frac{j^2}{\alpha^2}- \frac{sj}{\alpha}  + \frac{1}{4} \right)}}(Kl) + A_2 J_{-\sqrt{\left( \frac{j^2}{\alpha^2}- \frac{sj}{\alpha}  + \frac{1}{4} \right)}}(Kl) ,
\label{psiA.21}
\end{equation}
\begin{equation}
\psi_{B}= B_1 J_{\sqrt{\left( \frac{j^2}{\alpha^2}+ \frac{sj}{\alpha}  + \frac{1}{4} \right)}}(Kl) + B_2 J_{-\sqrt{\left( \frac{j^2}{\alpha^2}+ \frac{sj}{\alpha}  + \frac{1}{4} \right)}}(Kl) .
\label{psiB.21}
\end{equation}

Consider now the case of $j \rightarrow 0$. In this case only part of the solutions (\ref{psiA.2}) and (\ref{psiB.2}) are obtained.The cosine terms could be obtained by the limit $j \rightarrow 0$ of the function $J_{-\sqrt{\left( \frac{j^2}{\alpha^2} \pm \frac{sj}{\alpha}  + \frac{1}{4} \right)}}(Kl)$. In this case, the index of the Bessel function would be less than $-\frac{1}{2}$ and thus the wavefunction can could not be normalized. The limit is therefore
 \begin{equation}
\psi_{A}= \frac{1}{\sqrt{l}}A_1 \sin(Kl) ,
\label{psiAj->0}
\end{equation}
\begin{equation}
\psi_{B}= \frac{1}{\sqrt{l}}B_1 \sin(Kl) .\label{psiBj->0}
\end{equation}
The solution of the radial part of the wave function for $j=0$ is given by the relations (\ref{psiA.2}) and (\ref{psiB.2}). Notice that for the limit $j\rightarrow0 $, the cosine part of the solutions is lost, leading to (\ref{psiAj->0}) and (\ref{psiBj->0}). This abrupt change in the solution with a little change in the parameter $j$ was referred as a "scar" of the classical instability in Ref. \cite{kowalski2013dynamics}. As in the nonrelativistic case, the relativistic problem of a free particle on a double cone surface solved by the Dirac equation presents scars of the classical instability . Let us now investigate the case of a particle in the presence of a potential produced by an uniform magnetic field.

\section{Dirac equation of a double cone surface in the presence of an uniform magnetic field}
\label{4}

\begin{figure}[hpt]%[hbtp]
\centering
\includegraphics[width=9cm,height=8cm]{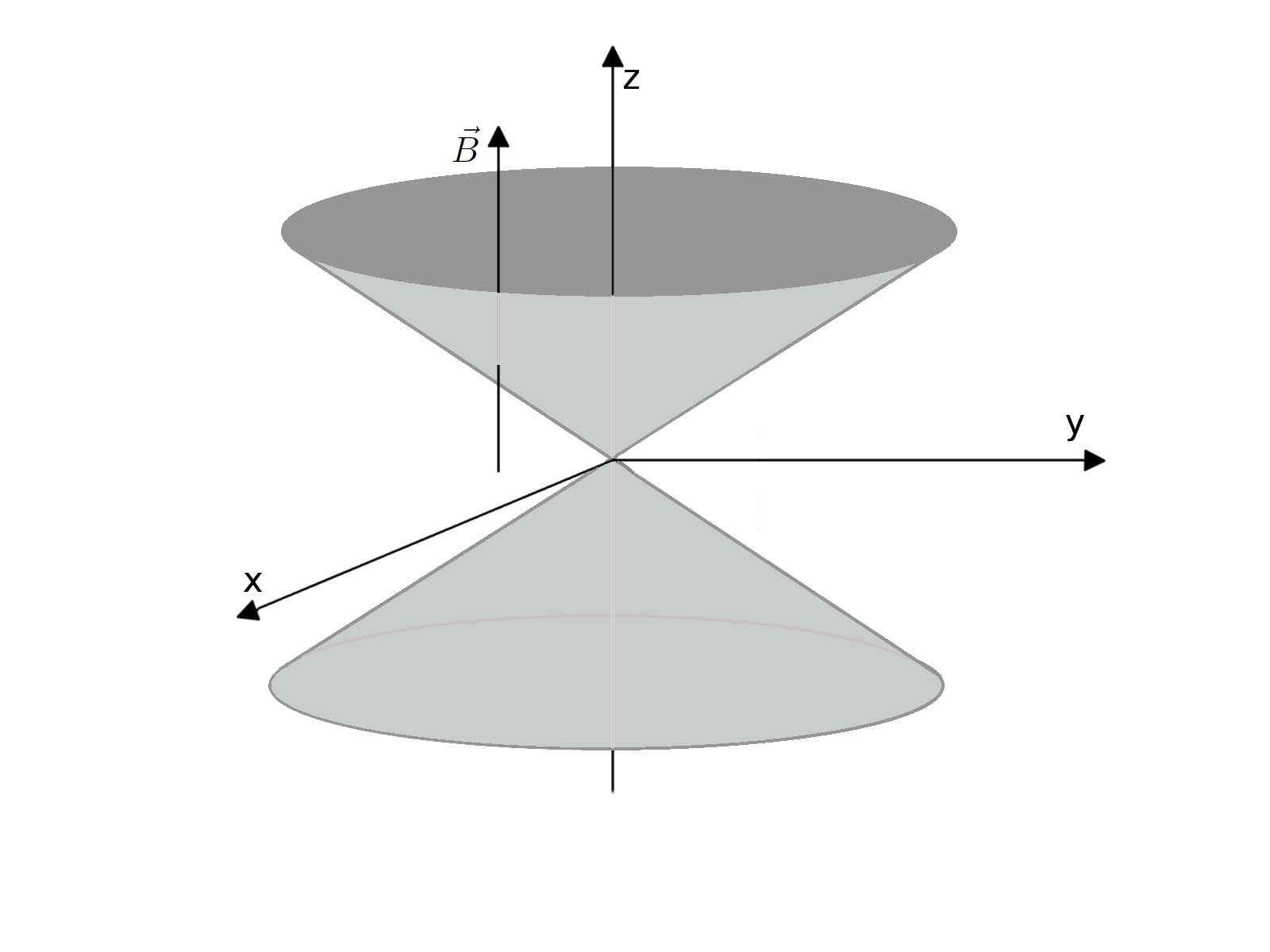}	
\caption{A double cone surface in the presence of a azimuthal magnetic field.}\label{figure2}
\end{figure}

We consider now the case of a particle on a double cone surface in the presence of an uniform magnetic field in the direction of the z-axis, $ \vec{B}=B_0 \hat{z}$, as can be seen in Fig. (\ref{figure2}).
The magnetic field is incorporated to the Dirac equation by minimal coupling. So, we have that
\begin{equation}
[i \hbar c \gamma^\mu (\partial _\mu + \Omega _\mu ) +e\gamma^\mu A_\mu- mc^2]\Psi=0 ,
\label{diracmagnetic}
\end{equation}
where $e$ is the electric charge. In the previous section, we determined the spinorial connection (\ref{spinorialconn}), as well as the gamma matrices (\ref{gammas}). The only quantity that we have to specify here is the vector potential $A_\mu$. The choice of the vector potential must take into consideration the metric of the surface, since it is related to the magnetic field by $\vec{\nabla} \times \vec{A}=\vec{B}$, and the curl is directly dependent on the metric of the surface. Taking into account the metric (\ref{metric}) and remembering that we extended the domain of the variable $l$, we obtain that
\begin{equation}
\vec{A}=\frac{B_0 \alpha  \left| \vec{l} \ \right| }{2} \hat{\phi} ,
\label{potencialvector}
\end{equation}
where the modulus of $\vec{l}$ appears here to ensure that the direction of $\vec{B}$ is preserved after extending the domain of $l$. With the results (\ref{spinorialconn}) and (\ref{potencialvector}) and defining the field locally, we can write the Dirac equation (\ref{diracmagnetic}) for the problem of a particle in a double cone surface in presence of a magnetic field as
\begin{eqnarray}
\fl i \hbar c \left[ \frac{\gamma^0}{c}\partial _t  + \gamma^1 \partial_l  + \frac{\gamma^2 }{\alpha l} \partial _\phi + \frac{\gamma^2}{\alpha l} \left(-\frac{i}{2} \alpha
\Sigma^3 \right) \right]\Psi \\\nonumber
+e \gamma^2 \left( \frac{B_0 \alpha  \left| \vec{l} \ \right| }{2} \right)\Psi -mc^2
 \Psi=0 .
 \label{magnetic1}
\end{eqnarray}
In order to solve this equation, we choose the \textit{ansatz} (\ref{ansatz}) and after some algebraic manipulations we obtain two coupled differential equations given by
\begin{eqnarray}
\fl \left(E- mc^2  \right)\psi_A  = -i\hbar c \sigma^1 \left(\partial _l + \frac{1}{2l}\right)\psi _B + \frac{\hbar c}{\alpha l} \sigma^2 j \psi _B \\\nonumber
- \left(\frac{eB_0\alpha \left| \vec{l} \ \right|}{2}\right) \sigma^2  \psi_B,
\label{mcoupled1}
\end{eqnarray}
and
\begin{eqnarray}
\fl \left(E+ mc^2 \right)\psi_B=-i\hbar c \sigma^1 \left(\partial _l + \frac{1}{2l}\right)\psi _A + \frac{\hbar c}{\alpha l} \sigma^2 j \psi _A \\\nonumber
- \left(\frac{eB_0\alpha \left| \vec{l} \ \right|}{2}\right) \sigma^2  \psi_A ,
\label{mcoupled2}
\end{eqnarray}
To uncouple the above equations and work with a unified solution, we define the quantities
\begin{equation}
\lambda = \left\{\begin{array}{cc}
+1, & \mbox{for} \ \psi_A \\
-1, & \mbox{for} \ \psi_B  \end{array} \right. ,
\label{lambda}
\end{equation}
and
\begin{equation}
\eta = \frac{\left| \vec{l} \ \right|}{l} =  \left\{\begin{array}{cc}
+1, & \mbox{for} \ l \ge 0 \\
-1, & \mbox{for} \ l < 0  \end{array} \right. .
\label{eta}
\end{equation}
It is important to note that the parameter $\eta$ is a new quantum number that indicates in which nappe of the double cone the particle is. With these quantities, and the parameter $s$ defined in the previous section, we can uncouple the differential equations (\ref{mcoupled1}) and (\ref{mcoupled2}) and obtain
\begin{eqnarray}
\frac{d^2}{dl^2} \psi_\lambda + \frac{1}{l} \frac{d}{dl} \psi_\lambda - \frac{1}{l^2} {M_\lambda}^2 \psi_\lambda + K_\lambda \psi_\lambda - \left(\frac{eB_0\alpha}{2\hbar c}\right)^2 l^2 \psi_\lambda = 0  ,
\label{solution1}
\end{eqnarray}
where
\begin{eqnarray}
{M_\lambda} = \left(\frac{j}{\alpha}-\frac{\lambda s}{2}\right) ,
\label{amount1}
\end{eqnarray}
and
\begin{eqnarray}
K_\lambda  = \left(\frac{E^2}{{\hbar}^2 c^2}-\frac{m^2 c^2}{\hbar^2} \right)
+ \frac{eB_0 \eta}{2\hbar c}\left( j+\lambda s \alpha\right)  .
\label{amount2}
\end{eqnarray}
In order to simplify the equation (\ref{solution1}) we perform the change of variables $ \zeta=\left(\frac{eB_0\alpha}{2\hbar c}\right)l^2$. With this new coordinate, we write
\begin{equation}
\frac{d^2}{d\zeta^2} \psi_\lambda + \frac{1}{\zeta} \frac{d}{d\zeta} \psi_\lambda - \frac{1}{4\zeta^2} {M_\lambda}^2 \psi_\lambda + \frac{K^{'}_\lambda}{\zeta} \psi_\lambda - \frac{1}{4} \psi_\lambda = 0,
\end{equation}
where
\begin{equation}
K^{'}_\lambda = \frac{\hbar c }{2 e B_0 \alpha}K_\lambda.
\end{equation}
To solve this equation, we look at its asymptotic behavior  in the limits $l \rightarrow \pm\infty$ and $l \rightarrow 0$. This technique involves the evaluation of the solutions in the asymptotic limits and the proposal of a function which gives the same solution at these limits. By using this approach, we obtain the solution
\begin{eqnarray}
\psi_\lambda = e^{-\frac{\zeta}{2}} \zeta^{\frac{\left|M_\lambda \right|}{2}}F(\zeta) .
\label{assintotic}
\end{eqnarray}
Replacing the solution (\ref{assintotic}) into Eq. (\ref{solution1}), we get the following equation
\begin{eqnarray}
\zeta \frac{d^2}{d\zeta ^2} F(\zeta) + \left( \left| M_\lambda \right| + 1 - \zeta \right) \frac{d}{d\zeta} F(\zeta) + \left( K^{'}_\lambda - \frac{\left| M_\lambda \right| + 1}{2} \right)F(\zeta) = 0 .
\label{Hypergeometric}
\end{eqnarray}
This is a confluent hypergeometric equation and $F(\zeta)$ is a confluent hypergeometric function that can be expressed as $ F \left( \frac{\left| M_\lambda \right| +1}{2} - K^{'}_\lambda , \left| M_\lambda \right| +1 ; \zeta \right) $. With this solution, we write the eigenfunction as
\begin{eqnarray}
\fl \Psi (t, l, \phi) = e^{\frac{-iEt}{\hbar}} e^{i j \phi} \left( \frac{eB_0 \alpha}{2 \hbar c}\right)^{\frac{\left| M_\lambda \right|}{2}}
e^{-\left( \frac{eB_0 \alpha}{4 \hbar c} \right) l^2} \left| \vec{l} \ \right|^{\left| M_\lambda \right|} \times \\\nonumber
 F \left( -n, \left| M_\lambda \right| +1 ;\frac{eB_0 \alpha}{2 \hbar c} l^2 \right) .
\label{eigenfunction}
\end{eqnarray}

\begin{figure}[h]
\center
\subfigure{\includegraphics[width=7cm]{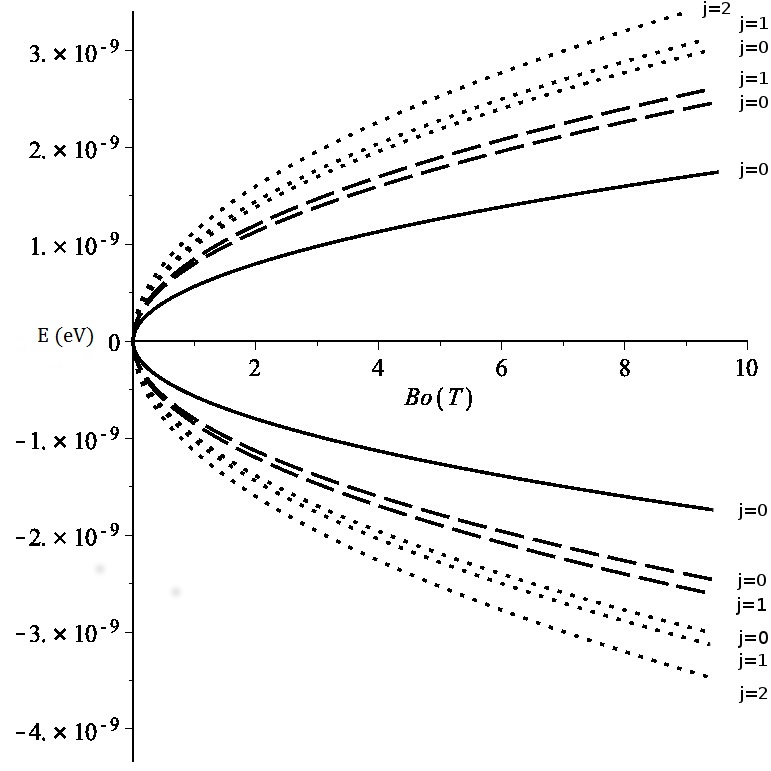}}
\qquad
\subfigure{\includegraphics[width=7cm]{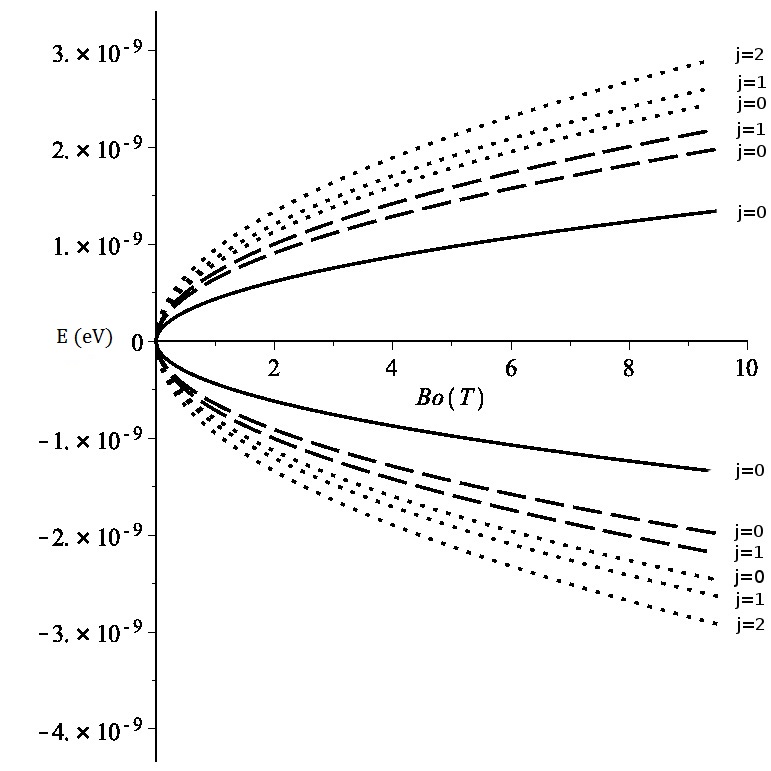}}
\caption{Energy levels for a double cone surface in the presence of an uniform magnetic field parallel to its axis. We considered here $\lambda=s=\hbar=c=+1$ and $\eta=-1$. Different values of $n$ and $j$ were considered: $n=0$ (solid line), $n=1$ (dashed line) and $n=2$ (dotted line). For comparison, in (a)  we considered the planar case, that is, $\alpha=1$; in (b) a double cone with $\alpha=0.7$. }
\label{EvsB}
\end{figure}

To get physically acceptable solutions, that is, normalizable solutions, we need to truncate the hypergeometric series. This condition lead us to impose that $-n= \left(-K^{'}_\lambda + \frac{\left|M_\lambda \right| +1} {2} \right)$, where $n \in Z$. From this condition, we obtain that the energy spectrum is given by
\begin{eqnarray}
E^2 = 2eB_0 \hbar  c \alpha \left\{ n + \frac{1}{2} + \frac{1}{2} \left| \frac{j}{\alpha} -\frac{\lambda s}{2} \right|  - \frac{\eta }{4 \alpha} \left( j + \lambda s \alpha \right) \right\} + m^2c^4.
\label{energyspectrum}
\end{eqnarray}
As expected, the energy spectrum depends on the parameter $\alpha$, which gives a measure of the opening angle of the conical surface. As discussed earlier, we considered here that this parameter assumes values in the interval $ 0 < \alpha < 1 $, which is the interval where we have a double cone surface. In this interval, we notice that the energy levels in the case of a conical surface are smaller than the corresponding ones in the planar case ($\alpha = 1$). It is possible to see the decreasing of the energy due to the parameter $\alpha$ in Fig. \ref{EvsB} , where we plot some states with $\alpha=1$ (Figure 3 (a)) and $\alpha = 0.7$ (Figure 3 (b)). 

A curious behavior of the spectrum (\ref{energyspectrum}) is noticed when we observe the influence of the parameter $\eta$. As we established in (\ref{eta}), this parameter depends on which cone the particle is. Therefore, the energies of particles in a double cone surface in the presence of a magnetic field are different for each cone, considering the same values for the quantum numbers $n$ and $j$. This comes from the coupling of the magnetic field to the total angular momentum as seen in Eq. (\ref{energyspectrum}).

\section{Concluding Remarks}
\label{5}

In this work, we study the quantum relativistic problem of a free particle constrained to a double cone surface and a particle with and without the presence of an uniform magnetic field. To work with the double cone surface, we build an appropriate coordinate system, since the usual spherical coordinates lead to some complications on the double cone surface. With this approach, we constructed the Dirac equation for a free particle and for a particle in presence of an uniform magnetic field. 

We expected to find instability in the motion on the relativistic case of a free particle and this behavior is indeed found. For $j=0$ the solution is a combination of sines and cosines, but for $j \rightarrow 0$, the cosine terms is lost due the closure relation for the Bessel functions. We see then that the radial solution for the wave function changes abruptly around $j \rightarrow 0$, as observed on the radial wave function in the nonrelativistic problem. In the problem with a magnetic field, the geometry of the surface leads to a curious behavior of the energy spectrum, it shows indirect dependence with the radial coordinate, presenting a different spectrum for each nappe. Since the electronic structure of materials like graphene and topological insulators can be described by an effective Dirac equation, it would be interesting to investigate, how the magnetic field affects the energy spectrum of a graphene double cone, which could be useful in future applications of graphene-based devices. We expect to address this problem in forthcoming publications.

\ack

 This work was supported by the CNPq, CAPES, FAPEMA and FACEPE (Brazilian agencies).

\section*{References}
\bibliographystyle{iopart-num}
\bibliography{ref}

\providecommand{\newblock}{}
\begin{thebibliography}{10}
\expandafter\ifx\csname url\endcsname\relax
  \def\url#1{{\tt #1}}\fi
\expandafter\ifx\csname urlprefix\endcsname\relax\def\urlprefix{URL }\fi
\providecommand{\eprint}[2][]{\url{#2}}

\bibitem{klitzing}
{Klitzing} K~V, {Dorda} G and {Pepper} M 1980 {\em Physical Review Letters\/}
  {\bf 45} 494--497

\bibitem{novoselov2004electric}
Novoselov K~S, Geim A~K, Morozov S, Jiang D, Zhang Y, Dubonos S, Grigorieva I
  and Firsov A 2004 {\em science\/} {\bf 306} 666--669

\bibitem{qi2010quantum}
Qi X~L and Zhang S~C 2010 {\em Physics Today\/} {\bf 63} 33--38

\bibitem{filgueiras2008quantum}
Filgueiras C and Moraes F 2008 {\em Annals of Physics\/} {\bf 323} 3150--3157

\bibitem{furtado1999landau}
Furtado C and Moraes F 1999 {\em EPL (Europhysics Letters)\/} {\bf 45} 279

\bibitem{bakke2010bound}
Bakke K and Furtado C 2010 {\em Physical Review D\/} {\bf 82} 084025

\bibitem{furtado1997electrostatic}
Furtado C and Moraes F 1997 {\em Classical and Quantum Gravity\/} {\bf 14} 3425

\bibitem{schmeltzer2012geometrical}
Schmeltzer D 2012 {\em New Journal of Physics\/} {\bf 14} 063025

\bibitem{ran2009one}
Ran Y, Zhang Y and Vishwanath A 2009 {\em Nature Physics\/} {\bf 5} 298--303

\bibitem{lima2014effects}
Lima J~R, Brand{\~a}o J, Cunha M~M and Moraes F 2014 {\em The European Physical
  Journal D\/} {\bf 68} 1--7

\bibitem{bueno2012landau}
Bueno M, Furtado C and de~M~Carvalho A 2012 {\em The European Physical Journal
  B-Condensed Matter and Complex Systems\/} {\bf 85} 1--5

\bibitem{kowalski2013dynamics}
Kowalski K and Rembieli{\'n}ski J 2013 {\em Annals of Physics\/} {\bf 329}
  146--157

\bibitem{lopes2015theoretical}
Lopes M~D, Azevedo S, Moraes F and Machado M 2015 {\em The European Physical
  Journal B\/} {\bf 88} 1--6

\bibitem{jonasjap}
Lima J~R~F 2015 {\em Journal of Applied Physics\/} {\bf 117} 084303

\bibitem{vozmediano2010gauge}
Vozmediano M~A, Katsnelson M and Guinea F 2010 {\em Physics Reports\/} {\bf
  496} 109--148

\bibitem{birrell1984quantum}
Birrell N and Davies P 1984 {\em Quantum Fields in Curved Space\/} Cambridge
  Monographs on Mathematical Physics (Cambridge University Press) ISBN
  9780521278584
  \urlprefix\url{https://books.google.com.br/books?id=SEnaUnrqzrUC}

\bibitem{misner1973gravitation}
Misner C, Thorne K and Wheeler J 1973 {\em Gravitation\/} ({\em Gravitation\/}
  no pt. 3) (W. H. Freeman) ISBN 9780716703440
  \urlprefix\url{https://books.google.com.br/books?id=w4Gigq3tY1kC}

\end{thebibliography}

\end{document}